\documentclass[reprint,amsmath,amssymb,prl, aps, superscriptaddress]{revtex4-2}

\usepackage{graphicx}
\usepackage{dcolumn}
\usepackage{bm}
\usepackage{physics}
\usepackage[unicode=true,pdfusetitle,bookmarks=false,breaklinks=false,pdfborder={0 0 0},backref=false,colorlinks=false]{hyperref}

\usepackage[english]{babel}
\makeatletter
\let\ORIbbl@fixname\bbl@fixname
\def\bbl@fixname#1{%
  \@ifundefined{languagealias@\expandafter\string#1}
    {\ORIbbl@fixname#1}
    {\edef\languagename{\@nameuse{languagealias@#1}}}%
}
\newcommand{\definelanguagealias}[2]{%
  \@namedef{languagealias@#1}{#2}%
}
\makeatother
\definelanguagealias{en}{english}

\begin{document}

\title{Pulsed to continuous-wave quantum dot cavity-QED
}

\author{Mio Poortvliet}
\email{poortvliet@physics.leidenuniv.nl}
\affiliation{%
Huygens-Kamerlingh Onnes Laboratory, Leiden Institute of Physics, Leiden University, The Netherlands
}
\author{Petr Steindl}
\affiliation{%
Huygens-Kamerlingh Onnes Laboratory, Leiden Institute of Physics, Leiden University, The Netherlands
}

\affiliation{%
Universit\'e Paris-Saclay, Centre de Nanosciences et de Nanotechnologies, CNRS, 10 Boulevard Thomas Gobert, 91120 Palaiseau, France
}
\author{Wolfgang L\"offler}
\affiliation{%
Huygens-Kamerlingh Onnes Laboratory, Leiden Institute of Physics, Leiden University, The Netherlands
}

\date{\today}

\begin{abstract}
Resonant cavity--quantum system interactions are well understood in the pulsed coherent and continuous-wave steady-state limits, but not much in between. We investigate the intermediate regime, where pulse duration, coupling, decoherences, and decays occur on comparable timescales, and pulse bandwidth matches cavity splitting and detunings. We reveal an interplay of cavity-enhanced excitation and Purcell-enhanced emission leading to warped chevron patterns. We develop a quantum master-equation model that efficiently treats the excitation-light coherent states in photon number space and reproduces the experimental data. Using these results we identify how polarization-split cavities can optimize single-photon purity and emission probability.
\end{abstract}

\maketitle

Self-assembled InGaAs quantum dots (QDs) in optical micro cavities are a promising source of bright quantum states of light \cite{steindlArtificialCoherentStates2021, snijdersPurificationSinglephotonNonlinearity2016, trivediGenerationNonClassicalLight2020, meekhofGenerationNonclassicalMotional1996}, such as single \cite{somaschiNearoptimalSinglephotonSources2016, tommBrightFastSource2021}, bunched \cite{munozEmittersNphotonBundles2014, fischerPulsedRabiOscillations2017, fischerSignaturesTwophotonPulses2017}, and entangled \cite{gaoObservationEntanglementQuantum2012, huetDeterministicReconfigurableGraph2025, costeHighrateEntanglementSemiconductor2023, mullerOndemandGenerationIndistinguishable2014} photons. Resonant excitation enables QD spin control \cite{hogg_fast_2025, huet_dynamical_2026}, nuclear spin cooling \cite{nguyenEnhancedElectronSpinCoherence2023, koong_coherent_2026} and record brightness single-photon emission \cite{wang_towards_2019, huetIndustryreadySpinphotonInterfaces2026, tommBrightFastSource2021, dingHighefficiencySinglephotonSource2025}, and is well studied in two limiting timescales \cite{schofieldPhotonIndistinguishabilityMeasurements2022} characterized by distinct features: In the pulsed regime, Rabi oscillations are observed \cite{rabiSpaceQuantizationGyrating1937, jaynesComparisonQuantumSemiclassical1963, gentileExperimentalStudyOne1989}, enabling deterministic population inversion and coherent control of quantum light by the excitation pulse properties \cite{flaggResonantlyDrivenCoherent2009, stievaterRabiOscillationsExcitons2001, hanschkeExperimentalMeasurementReappearance2024, schaibleyDirectDetectionTimeresolved2013}. In the steady-state regime, the Mollow triplet appears under continuous-wave excitation \cite{mollowPowerSpectrumLight1969, flaggResonantlyDrivenCoherent2009, moelbjergResonanceFluorescenceSemiconductor2012}, where now the system response is fully captured in the frequency domain. In the intermediate regime where pulse durations are comparable to the cavity-enhanced QD lifetime, the time-bandwidth uncertainty means the excitation is neither temporally nor spectrally well-defined, and understanding the dynamics requires merging these two pictures \cite{wiggerResonancefluorescenceSpectralDynamics2021}. This regime probes the transition where coherent control breaks down due to dephasing and decoherence, but the steady-state solution has not yet been reached \cite{moelbjergResonanceFluorescenceSemiconductor2012}. 

The dynamics of Rabi oscillations are governed by the pulse area
\begin{equation}
    \Theta=\frac{1}{\hbar}\int_T \vec \mu \cdot \vec E(t)\mathrm{d}t,\label{eq:pulse_area}
\end{equation}
where $T$ is a time range containing the pulse, $\vec \mu$ is the electric dipole moment of the QD transition \cite{fischerPulsedRabiOscillations2017, hanschkeQuantumDotSinglephoton2018}, and crucially $\vec E$ is the envelope of the electric field -- related to the intensity as $E\propto\sqrt{I}$.
Typically, experimental studies of pulsed resonance fluorescence have varied the pulse area by changing the laser intensity at a fixed pulse duration \cite{stievaterRabiOscillationsExcitons2001, javadiCavityenhancedExcitationQuantum2023}. Equation \ref{eq:pulse_area} states that the same pulse area can be obtained using lower intensity light in a longer pulse, which keeps phonon coupling weak \cite{nazirModellingExcitonPhonon2016, reiterRolePhononsExciton2014} and requires less photons per pulse, relaxing the requirements on the suppression of the excitation laser \cite{gonzalez-ruizTwophotonCorrelationsHOM2025}.
Only a few studies use the pulse duration as the control parameter, and they focused on multi-photon emission \cite{hanschkeQuantumDotSinglephoton2018, fischerPulsedRabiOscillations2017, fischerSignaturesTwophotonPulses2017}. 
Here we vary the pulse duration and focus on the cavity-QD dynamics: we explore the crossover from pulsed to continuous-wave dynamics in both the time and frequency domains by changing the laser pulse duration and the detunings between the laser, the QD transition, and the cavity modes. We identify unexpected parameter regimes for maximal photon extraction and purity given by the interplay of dynamics and detuning of the driving pulse, cavity resonance and QD energy.

\begin{figure}
    \centering
    \includegraphics[width=\linewidth]{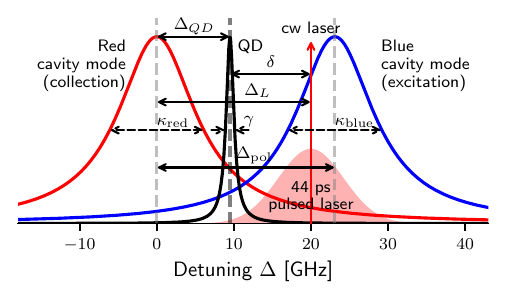}
    \caption{Schematic overview of the relevant detunings, showing the polarization-split cavity modes, QD transition and the scanning laser in both continuous-wave and pulsed mode. The red (blue) cavity mode is labeled as the collection (excitation) mode. All detunings are defined with respect to the collection mode frequency, consistent with the convention throughout this work.}
    \label{fig:spectral_configuration}
\end{figure}

Epitaxially grown self-assembled QDs intermediately coupled to an optical microcavity are particularly well suited to explore this crossover. The Purcell effect shortens the radiative lifetime and broadens the linewidth, placing the system in a regime where the cavity loss rate $\kappa$, the Purcell-enhanced QD decay rate $\gamma$, and the cavity-QD coupling rate $g$ are within one order of magnitude at around 1-10 GHz
, making both the temporal and spectral features experimentally resolvable by spectral tuning of the QD and laser, and tunable-duration laser pulses. For the latter, we use programmable pulses with durations from 17 ps to 1 ns \cite{poortvlietPicosecondLaserPulses2025}, corresponding to a spectral bandwidth of up to 25 GHz, covering all energy scales of the QD-cavity system, see Fig. \ref{fig:spectral_configuration}. 

We record the rate of fluorescence photons and simultaneously measure the second-order correlation function $\mathrm{g}^{(2)}(0)$, which shows a non-trivial pattern of photon bunching and anti-bunching across the pulsed to continuous-wave crossover. The observed photon bunching is consistent with the emission of photon bundles, as observed for resonantly driven two-level systems \cite{fischerSignaturesTwophotonPulses2017}, while the anti-bunching reflects the single-photon character of the emitter. We first develop a simple analytic theory describing the generalised Rabi frequency as a function of laser, QD, and cavity detunings, which agrees well to the experimental data. We then develop a quantum master-equation model that fully captures the intermediate coupling regime and also quantitatively reproduces the experimentally observed dynamics. To our knowledge, this is the first model describing time-dependent pulsed driving in an intermediately coupled polarization-split cavity-QD system, and we make the implementation openly available to support future work in this regime \cite{poortvliet_pulsed_2026}.

Experimentally, we use a single self-assembled singly-charged InGaAs QD in an oxide-apertured optical micro-cavity (details in Ref. \cite{steindlResonantTwoLaserSpinState2023}) with a controlled frequency splitting between the two orthogonally polarized fundamental cavity modes \cite{bakkerPolarizationDegenerateMicropillars2014}, which we refer to as the red and blue modes. The cavity and QD are designed to operate near a wavelength of $935$ nm. The QD is embedded in a p-i-n diode structure, allowing us to deterministically charge it and electrically fine-tune its transition frequency between the two cavity resonances via the quantum confined Stark effect. We emphasize that, since our QD is singly charged and the optical trion transitions are energy degenerate, the total QD -- cavity coupling does not depend on polarization. The quantum confined Stark effect-tuning allows to choose the red (blue) or blue (red) cavity mode for collection (excitation), revealing any phonon effects \cite{javadiCavityenhancedExcitationQuantum2023}.

We first characterize the device at 4\,K using a continuous-wave tunable narrow-linewidth diode laser scanning across the cavity resonance. We measure the reflected light of the cavity-QD system as a function of laser excitation frequency and polarization, and fit the data to a semiclassical model (see Supplemental Material \ref{app:semi_classical_model}) \cite{snijdersExtendedPolarizedSemiclassical2020} to extract all system parameters. We obtain a cavity linewidth $\kappa/2\pi=12$ GHz and polarization mode splitting $\Delta_\mathrm{pol}=f_\mathrm{blue}-f_\mathrm{red}=22$ GHz, a bare QD linewidth $\gamma_0/2\pi=0.24$ GHz (pure dephasing is negligible) and QD-cavity coupling rate $g/2\pi\approx3$ GHz, corresponding to a cooperativity $C\approx3$ resulting in a moderate Purcell enhancement $F_p=1+4C\approx13$. Our device with $\Delta_\mathrm{pol}/\kappa \approx 2$ is predicted to function as an efficient resonance fluorescence single-photon source \cite{wang_towards_2019} and operates near the `Goldilocks condition' $\kappa \approx 2g$ \cite{esmann_solid-state_2024}, therefore the understanding of the internal system dynamics is of high relevance. However, as mentioned above, $\kappa, \gamma, g$ and the transform-limited laser pulse bandwidth are all comparable, such that no parameter dominates the dynamics and no simplification can be applied.

To study the pulsed to continuous-wave crossover, we excite the QD near-resonance by laser pulses made by gating continuous-wave narrow-linewidth laser light using two cascaded high-bandwidth electro-optic modulators (EOMs) driven by a custom electronic pulse compressor \cite{poortvlietPicosecondLaserPulses2025}, allowing control over the interaction time between the driving field and the QD-cavity system. We detect resonance fluorescence using a confocal microscope in reflection, separating the excitation and collection modes using a cross-polarization technique \cite{steindlCrossPolarizationExtinctionEnhancementSpinOrbit2023}: in both the excitation and detection paths, the polarization is controlled by a combination of polarizer, a half wave and a quarter wave plate, and spatial filtering is done with single-mode fibers. 
We characterize the reflected light using a fiber-based Hanbury-Brown and Twiss detection setup with two single-photon avalanche detectors (350 ps jitter). We bin the coincidence counts in a 400 ps window and normalize to get the pulsed-mode second order autocorrelation function $\mathrm{G}^{(2)}(\tau_k)$ for each detection bin at $\tau_k$ using a procedure described in Ref. \cite{dadaIndistinguishableSinglePhotons2016} and Supplemental Material \ref{app:geetwo}.

We vary the laser pulse duration for different combinations of QD and laser detunings. As shown in Figure \ref{fig:spectral_configuration}, 
we define most detunings with respect to the collection cavity mode in terms of natural frequencies ($f=2\pi\omega$, $\Delta_{QD}=f_{QD}-f_{C}$ and $\Delta_{L}=f_L-f_{C}$) except for the QD-laser detuning $\delta$; for details on the calibration see the Supplemental Material Section \ref{app:calibration_of_detunings}.

We note that the polarization non-degenerate microcavity shows a strong frequency-dependent birefringence. This makes broadband optimal laser extinction impossible over a frequency range of more than a GHz, limiting the cross-polarization extinction ratio under pulsed excitation. Therefore, using longer pulses increases the cross-polarization contrast in two ways: First, the reduced spectral width of the pulse fits better in the available cross-polarization bandwidth. We measure up to two orders of magnitude difference in contrast between pulsed and continuous-wave excitation. Second and more generally, for a longer pulse, a much lower intensity can be used for population inversion -- Equation \ref{eq:pulse_area} shows that for fixed pulse area and shape, the intensity scales inversely with the square of the pulse length. 
For photon count rate measurements, we can correct for the laser leakage by subtracting the counts recorded when the QD is detuned far away from the cavity \cite{zhanDynamicalAcousticControl2025}, we refer to this as the `differential count rate'. A detailed explanation of our experimental protocol is described in Supplemental Material Section \ref{app:experimental_protocol}. 

\begin{figure}
    \centering
    \includegraphics{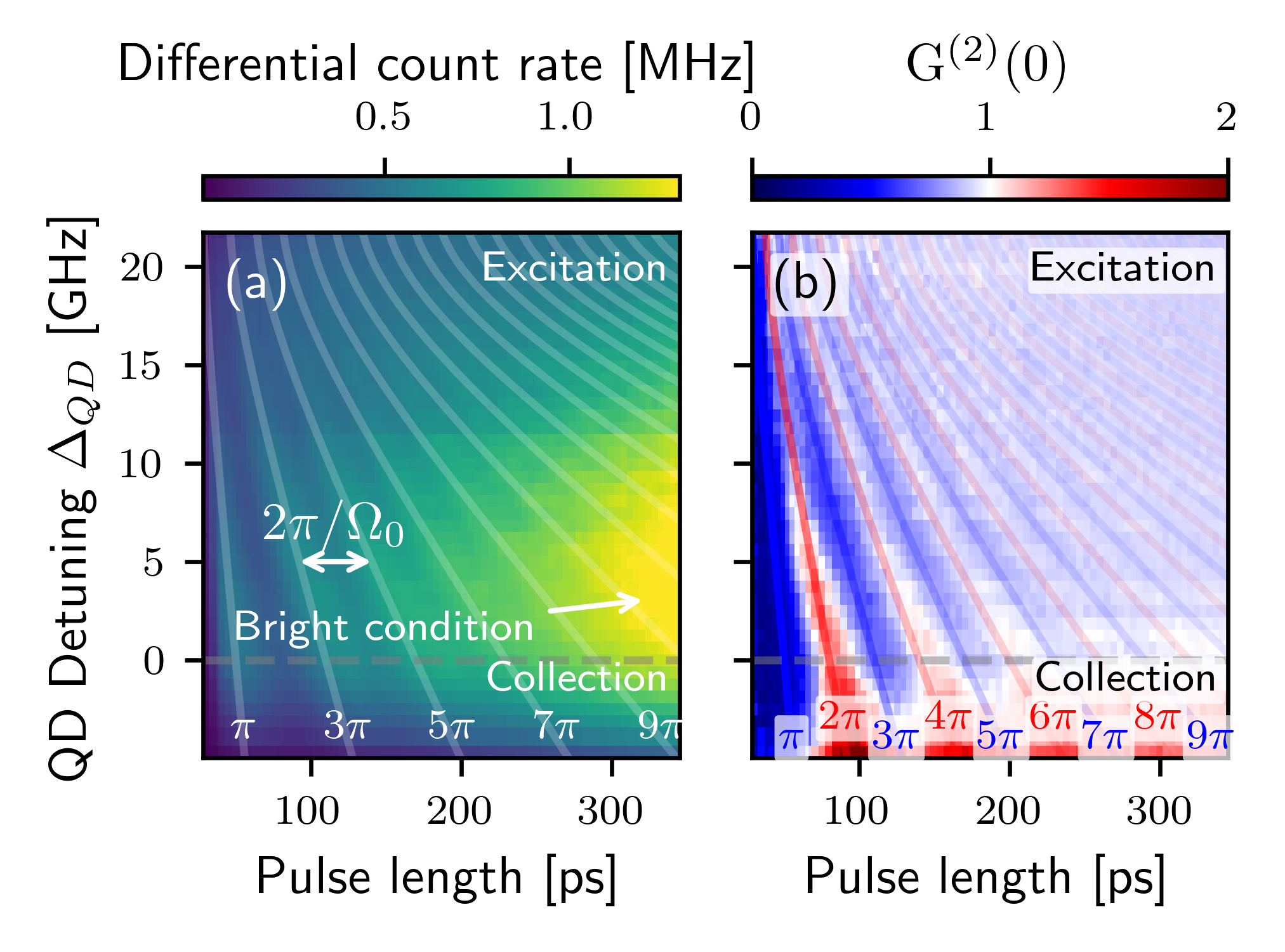}
    \caption{Resonance fluorescence as a function of QD-cavity detuning $\Delta_{QD}$ and pulse width, with the laser kept resonant with the QD transition ($\Delta_L = \Delta_{QD}$). The cavity mode frequencies are indicated by dashed vertical lines. (a) Differential count rate. Rabi oscillations are visible at short pulse widths whose frequency increases as the QD approaches the excitation mode due to the cavity enhancement of the driving field. White lines indicate odd-$\pi$ pulse positions predicted by the cavity-enhanced Rabi frequency model. The bright condition is indicated, as further discussed in the text. (b) Simultaneously recorded $\mathrm{g}^{(2)}(0)$, showing anti-bunching at odd-$\pi$ pulse areas and photon bundles at even-$\pi$ pulse areas. 
}
    \label{fig:cavity-QD-detuning-red}
\end{figure}

We first investigate the case that the laser is resonant with the QD ($\Delta_L=\Delta_{QD}$), for different detunings from the cavity modes. In this case, we expect that the QD undergoes cavity-modified Rabi oscillations at the Rabi frequency $\Omega_0=\frac{g|d|\hbar}{2\mathcal{E}_0}|E_\mathrm{cav}(\Delta_L)|,$ where $d=1/\sqrt{2}$ is the polarization overlap of the QD emission and the cavity mode \cite{snijdersExtendedPolarizedSemiclassical2020}, 
$\mathcal{E}_0$ is the root mean square vacuum electric field amplitude \cite{foxQuantumOpticsIntroduction2006}, and the intra-cavity driving field is
\begin{equation}
    E_\mathrm{cav}(\Delta_L) = \mathcal{E}_0 \frac{\alpha_0}{\kappa/2-i\,2\pi\left( \Delta_L + \Delta_\mathrm{pol} \right)} .\label{eq:cavity_enhanced_field}
\end{equation}
Here, $\alpha_0$ is the driving rate of the coherent laser light driving in the cavity (after the first mirror), calculated from the laser power $P_L$ and frequency $\omega_L$ as $\alpha_0 = \sqrt{P_{L}\kappa/\hbar \omega_L}$. The driving field is thus strongest when the laser is near to the excitation cavity mode, making $\Omega_0$ strongly dependent on $\Delta_{QD}$ around the excitation mode frequency. 

This is exactly what we observe in the experimental data in Figure \ref{fig:cavity-QD-detuning-red}(a), where we use an optimization routine described in Supplemental Material Section \ref{app:experimental_protocol} 
to keep the laser resonant with the QD.
At short pulse widths and small detunings, Fig. \ref{fig:cavity-QD-detuning-red}(a) shows clear Rabi oscillations whose visibility decreases at longer pulse lengths as population relaxation destroys the coherent dynamics. The white lines in Fig. \ref{fig:cavity-QD-detuning-red}(a) show the predicted odd-$\pi$ pulse positions using this model and the system parameters - we only use one overall intensity scaling parameter.

Figure \ref{fig:cavity-QD-detuning-red}(b) shows the simultaneously recorded zero-time pulsed mode second-order correlation function $\mathrm{G}^{(2)}(0)$. At short pulse widths the photon statistics oscillate between anti-bunching at odd-$\pi$ pulse areas, and bunching at even-$\pi$ pulse areas. The bunching reflects the emission of photon bundles: after a $2\pi$ pulse the QD returns to its ground state coherently, the residual fluorescence from re-excitation is dominantly multi-photon emission \cite{fischerPulsedRabiOscillations2017, fischerSignaturesTwophotonPulses2017}. As the pulse width increases $\mathrm{G}^{(2)}(0)$ tends to unity because the random emission events within a pulse become uncorrelated once population relaxation dominates over coherent dynamics. 
At detunings away from both cavity modes ($\Delta_{QD}<-3$ GHz) we measure apparent photon bunching due to our EOM-based optical pulses, as explained in Supplemental Material Section \ref{app:bunching}.

Using long pulse lengths for excitation the dynamics approach the continuous-wave regime. Here the system response varies little in the temporal domain, but in the frequency domain a parameter condition with maximal differential count rate emerges, indicated in Fig. \ref{fig:cavity-QD-detuning-red}(a). The position of this bright condition is determined by the competition between excitation enhancement (overlap laser - excitation mode), and QD Purcell enhancement (overlap QD - collection mode).

\begin{figure*}
    \centering
    \includegraphics{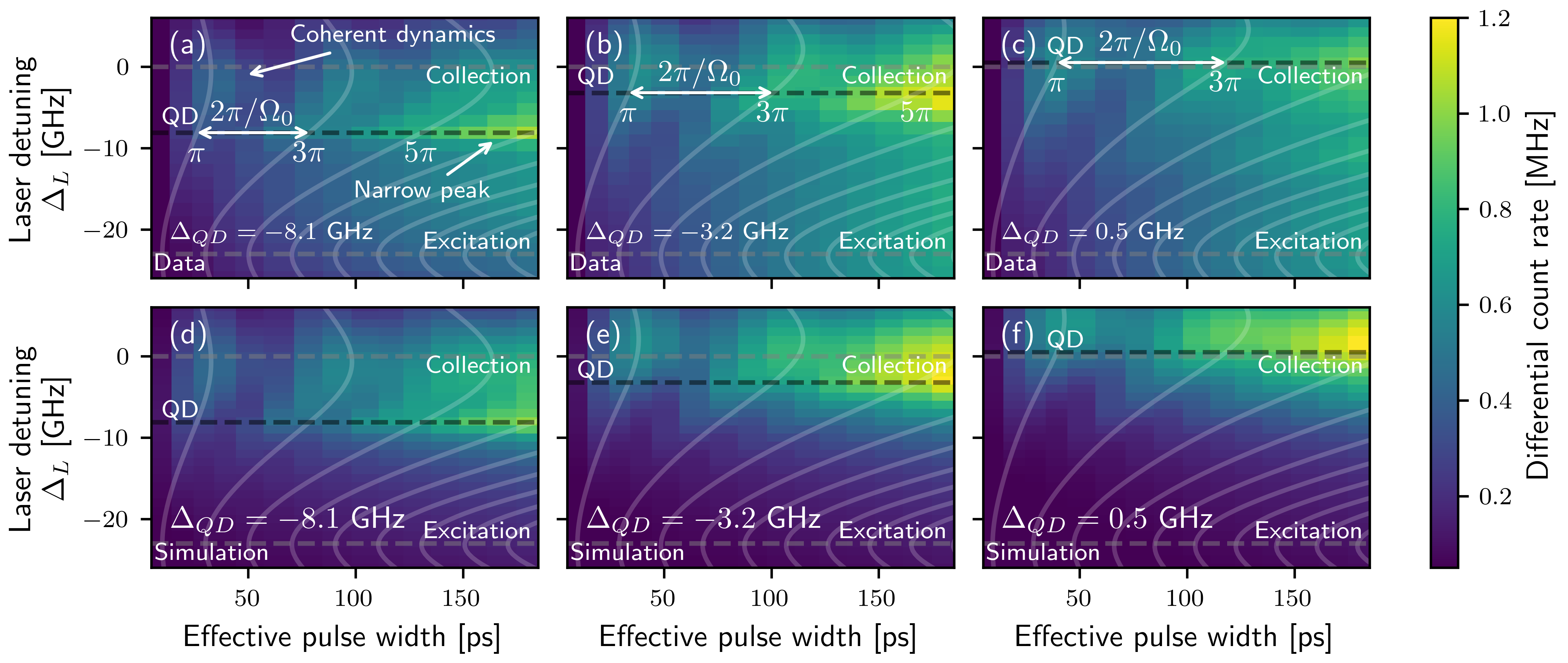}
    \caption{Near-resonant QD-cavity dynamics as a function of laser detuning $\Delta_L$ for three QD-cavity detunings $\Delta_{QD}$ (left to right). Dashed lines indicate the cavity mode frequencies and QD transition. Top row (a--c): differential count rate. Cavity enhancement of the driving field deforms the chevron pattern of detuned Rabi oscillations into a warped chevron, shifting the minimum Rabi frequency away from the QD resonance. White lines show odd-$\pi$ pulse positions from the model. Bottom row (d--f): quantum master equation simulations of panels (a--c), using system parameters extracted from a semiclassical model fit and a drive strength $\alpha_0 / 2\pi= 60.5$ GHz. 
    The "narrow peak" labeled in panel (a) is QD resonance fluorescence excited by approximately continuous-wave laser light due to EOM leakage.
    The simulations quantitatively reproduce all main features of the data.
    }
    \label{fig:cavity-laser-detuning}
\end{figure*}

We now investigate if the observed position of the bright condition can be explained by theory. We have to distinguish two regimes:
In the saturated regime ($\Omega_0 \gg \gamma$), the emission rate is limited by the QD lifetime $1/\gamma$, which is shortened by Purcell enhancement from both cavity modes:
\begin{equation}
\gamma(\Delta_{QD})/\gamma_0 \propto 1+D_\mathrm{col}(\Delta_{QD}) + D_\mathrm{ex}(\Delta_{QD}).
\end{equation}
Here the density of states $D_{\mathrm{col}(\mathrm{ex})}$ in the collection (excitation) mode is proportional to the intensity enhancement of a cavity, as given in Eq. \ref{eq:cavity_enhanced_field}, by $D_\mathrm{col}(\Delta_{QD})\propto\left|E_\mathrm{cav}(\Delta_{QD}-\Delta_\mathrm{pol})\right|^2$ and $D_\mathrm{ex}(\Delta_{QD})\propto\left|E_\mathrm{cav}(\Delta_{QD})\right|^2$.
The detected rate $R_\mathrm{det}^\mathrm{sat}$ is additionally increased by the efficient collection into the collection mode, which is proportional to the density of states:
\begin{equation}
    R_\mathrm{det}^\mathrm{sat}(\Delta_{QD}) \propto
     D_\mathrm{col}(\Delta_{QD}) \gamma(\Delta_{QD}).
\end{equation}
This rate is simply maximized in our system when the QD emission is resonant with the collection mode, placing the detuning of the bright condition at $\Delta_{QD}=0$.

In the Heitler regime ($\Omega_0 \ll \gamma$) the rate of QD emission scales linearly with the excitation intensity. Now both the excitation rate and the collection enhancement into the collection mode scale with the density of states. The detected rate is
\begin{equation}
    R_\mathrm{det}^\text{Heitler}(\Delta_{QD}) \propto D_\mathrm{col}(\Delta_{QD}) D_\mathrm{ex}(\Delta_{QD}) \gamma(\Delta_{QD}).
\end{equation}
Excitation and detection trade off symmetrically, placing the bright-condition detuning between the two cavity modes at $\Delta_{QD}=\Delta_\mathrm{pol}/2$. 

In our experiments (Figure \ref{fig:cavity-QD-detuning-red}) we have $\Omega_0/(2\pi) \approx 3$ GHz, which is in between the two regimes since $\Omega_0\sim\gamma$. The bright condition is at $\Delta_\mathrm{pol}/4$, which is as expected in between the two predictions, as indicated by the arrow in Fig. \ref{fig:cavity-QD-detuning-red}(a). 


\medskip
Before turning to pulse-width dependent Rabi dynamics, we first explain our simulation method. Due to the intermediate nature of the system parameters,  the QD hybridizes with both the excitation and detection cavity modes. To accurately model this, we must go beyond the optical Bloch equations which describe a weakly coupled system to solving the dynamics using the Lindblad quantum master equation. This is computationally resource-intensive because the Hilbert space required to describe the strong coherent driving field is large. We offload the description of this classical field by unitary transformation of the Hamiltonian using a carefully chosen displacement operator that follows the cavity response of the driving field. This drives the QD directly and allows the Hilbert space of the driving mode to remain on the order of the two level system-induced fluctuations \cite{fischerPulsedCoherentDrive2018}. Details of this procedure are given in the Appendix, Quantum master model. 
We implement this quantum master equation model using QuTiP \cite{qutip5}.

\medskip
We now study how the pulse-length dependent Rabi dynamics depends on the detuning between the laser and the QD, by changing $\Delta_L$ and keeping $\Delta_{QD}$ fixed. Figure \ref{fig:cavity-laser-detuning} shows measured data (panels a-c) and quantum master simulations (panels d-f) for three values of $\Delta_{QD}$. 

We focus first on the coherent dynamics. The laser-QD detuning $\delta = \Delta_L - \Delta_{QD}$ gives rise to the generalised Rabi frequency $\Omega$ which is dependent on the driving field detuning $\delta$ as
\begin{equation}
    \Omega^2(\delta) = \Omega_0^2 + \delta^2. \label{eq:rabi}
\end{equation}
This equation produces the characteristic chevron pattern of detuned Rabi oscillations \cite{yonedaFastElectricalControl2014, javadiCavityenhancedExcitationQuantum2023, rabiSpaceQuantizationGyrating1937, gentileExperimentalStudyOne1989}
\footnote{The term 'chevron pattern' in the context of Rabi oscillations appeared around 2000 in circuit-QED experiments \cite{cooperObservationQuantumOscillations2004,mcdermottSimultaneousStateMeasurement2005}}. The double sided arrow in each panel marks the bare Rabi frequency at laser-QD resonance, connecting these measurements to Fig.\ \ref{fig:cavity-QD-detuning-red}: as the QD is detuned away from the excitation mode (panels a--c, left to right), the intracavity driving field weakens and $\Omega_0$ decreases.

Because $\Omega_0$ varies with laser frequency through cavity enhancement (Eq. \ref{eq:cavity_enhanced_field}), the chevron pattern (in panel (a) the first lobe is labeled `coherent dynamics') is deformed or, more specifically, warped: the minimum Rabi frequency -- normally at $\delta = 0$ -- shifts to a laser detuning where the cavity enhancement decreases it faster than the detuning term in Eq. \ref{eq:rabi} increases it. Figure \ref{fig:cavity-laser-detuning} shows this deformation clearly. As far as we know, this is the first observation of warped chevron patterns in QD-cavity QED. Recently, a similar effect has been found in circuit QED \cite{thery_observation_2024}.

Ignoring the narrow peak, the brightest coherent emission (Fig. \ref{fig:cavity-laser-detuning}(a), `coherent dynamics') is where the driving laser is detuned from the QD resonance towards the collection mode. This is not expected in the weak coupling limit: the QD emits in its linewidth regardless of excitation frequency \cite{wangCoherenceResonanceFluorescence2025}, and the Purcell enhancement for emission into the collection mode should therefore be independent of laser detuning. Figure \ref{fig:cavity-laser-detuning} contradicts this: the emission is brightest when the laser is on the collection mode side of the QD resonance, not the QD resonance itself. The driving field hybridizes with the QD transition, pulling the emission frequency toward the driving frequency and thereby modulating the Purcell enhancement for the emission into the collection mode. This is a signature of the intermediate coupling regime and is not captured by a weak-coupling model. The quantum master equation simulations in Fig. \ref{fig:cavity-laser-detuning}(d--f) reproduce this behavior.

To exclude that the observed effects are caused by phonon-induced asymmetry \cite{javadiCavityenhancedExcitationQuantum2023, dingHighefficiencySinglephotonSource2025}, we repeated the experiments with swapped roles of the cavity modes. We observe no such effects, which is consistent with the relatively low driving intensities used here, the phonon coupling remains weak throughout \cite{nazirModellingExcitonPhonon2016, reiterRolePhononsExciton2014}. The full complementary dataset is presented in the Appendix, Role of cavity modes.

\medskip
Finally, we show how careful adjustment of the detunings and laser pulse length can be used to increase the purity and rate of the emission from a resonantly excited single-photon source.
The ratio of single photons and pump strength is proportional to $\frac{\Omega_0^2}{\Omega^2} \sin^2\left(\frac{\Omega t}{2}\right) / \frac{\alpha_0^2 t}{\kappa}$ in the short-pulse coherent limit. Inserting Eq. \ref{eq:cavity_enhanced_field} provides insight to detuning:
\begin{gather}
    \frac{R_{sp}}{R_\text{drive}}
    \propto
    \frac{\kappa g^2d^2\hbar^2/4}{\left(\kappa/2\right)^2+4\pi^2\left( \Delta_L + \Delta_\mathrm{pol} \right)^2} \frac{t}{\tilde{\Theta}^2}\sin^2\left(\frac{\tilde{\Theta}}{2}\right),\label{eq:photon_ratio}
\end{gather}
where $\tilde{\Theta} = t\sqrt{\Omega_0^2 + (\Delta_{QD} - \Delta_L)^2}$ is the generalized pulse area. This equation shows that longer pulses $t$ with the same (generalized) pulse area are beneficial for the purity while excitation laser leakage is the dominant impurity, and not QD re-excitation \cite{gonzalez-ruizTwophotonCorrelationsHOM2025}.

Using the spectral degree of freedom, tuning $\Delta_L$ affects the cavity response (first fraction in Eq. \ref{eq:photon_ratio}) faster than $\tilde{\Theta}$ if $\Omega_0$ is sufficiently large. In that case, by detuning the drive, the cavity enhancement gain can be stronger than the decreased population inversion -- consequently, detuning the laser from the QD is beneficial for the purity. We predict the detuning should be towards the excitation cavity mode, contrary to reported experimental results \cite{dingHighefficiencySinglephotonSource2025,javadiCavityenhancedExcitationQuantum2023}. This apparent discrepancy is explained considering the pulse shape: the cited studies use short, spectrally broad pulses which are shaped by the cavity. In this case, detuning away from the excitation mode concentrates the pulse around the QD resonance \cite{javadiCavityenhancedExcitationQuantum2023}. Eq. \ref{eq:photon_ratio} neglects the spectral width of the pulse, thus is valid for pulses longer than $2\pi/\kappa$. The intermediate value theorem implies that there exists a combination of $t, \kappa, \Omega_0$ yielding the highest purity exactly on the QD resonance, allowing full population inversion.

In conclusion, by pulse-length dependent QD cavity-QED experiments with a polarization non-degenerate microcavity, we have explored an intriguing regime where the laser pulse length is comparable to all relevant time-scales of the system. Despite the complexity of this regime, we were able to explain all features that are observed while changing the QD detuning, laser detuning and laser pulse length. Such measurements show warped chevron patterns which we model as cavity-enhanced Rabi oscillations. We have developed a quantum master model with a displacement-operator method to reduce the computational complexity of modeling strong laser pulses with a polarization non-degenerate microcavity. We find excellent agreement between theory and experiments, and an unexpected ideal configuration for single-photon generation. Further, we believe that the insight gained into cavity enhancement across time scales will contribute to spin control \cite{hogg_fast_2025, sunCavityEnhancedOpticalReadout2018, huetIndustryreadySpinphotonInterfaces2026} and cavity-enhanced QD-spin-mediated processing of incoming light with intermediate pulse length \cite{sun_quantum_2016, sun_single-photon_2018, parker_diamond_2024}.

\medskip
\begin{acknowledgments}
The devices were fabricated in collaboration with J. Frey, J, Norman, J. Bowers, and D. Bouwmeester at the University of California, Santa Barbara, USA. We acknowledge funding from NWO (680.92.18.04), NWO/OCW (Quantum Software Consortium Nos. 024.003.037 and 024.003.037/3368, Quantum Limits No. SUMMIT.1.1016), from the Dutch Ministry of Economic Affairs (Quantum Delta NL), and from the European Union’s Horizon 2020 research and innovation program under Grant Agreement No. 862035 (QLUSTER).

\end{acknowledgments}

\section{Appendix}
\subsection{Quantum master model}\label{app:quantum_master_model}
An energy degenerate singly charged QD can be modeled as a four level system with resident spin states $\ket{\uparrow}, \ket{\downarrow}$ and excited trionic states $\ket{T_\uparrow}, \ket{T_\downarrow}$. The QD with transition energy $\omega_{qd}$ ($\hbar=1$) is coupled with rate $g$ to a cavity with two polarization modes $X$ and $Y$ at energy $\omega_X, \omega_Y$. The Hamiltonian is \cite{snijdersExtendedPolarizedSemiclassical2020, arnoldMacroscopicRotationPhoton2015, javadiCavityenhancedExcitationQuantum2023}
\begin{multline}
    H = \sum_{i=X,Y} \omega_i \hat a^\dagger_i \hat a_i
    + \sum_{s=\uparrow, \downarrow} \omega_{qd} \ket{T_s}\bra{T_s}\\
    + \sum_{i=X,Y}\sum_{s=\uparrow, \downarrow} g \, \hat E_i \cdot \hat d_s \left( \ket{T_s}\bra{s}a_i + \ket{s}\bra{T_s}a^\dagger_i \right), \label{eq:hamiltonian}
\end{multline}
where the spin state $s\in\{ \uparrow, \downarrow \}$ determines the handedness of the circularly polarized transition dipole $\vec d = (d_X, d_Y)$. The transition dipole obeys $|\vec d|=1$, such that the light-matter coupling strength is completely described by $g$.
We model decoherence in a Lindblad master equation framework by the collapse operators $\mathcal{L}\sqrt{\kappa_X} \hat a_X, \mathcal{L} \sqrt{\kappa_Y}\hat a_Y, \mathcal{L}\sqrt{\gamma_0} \hat \sigma_-$ where $\mathcal{L}$ is the standard superoperator for decoherence.

We can drive the system in the $Y$-mode with a classical field of strength $\alpha (t)$ using the driving term
\begin{equation}
    H_{drive} =  i \left(\alpha(t) \hat{a}_Y^\dagger - \alpha^*(t) \hat{a}_Y \right),
\end{equation}
where $\alpha(t)$ can model a static (continuous-wave) or time-dependent (pulsed) driving field. From here we move to a frame rotating with the laser by the standard procedure \cite{foxQuantumOpticsIntroduction2006, snijdersExtendedPolarizedSemiclassical2020}. 

A large part of the driving mode Fock space is used to describe a classical field. In order to make the computation faster we eliminate the part of the Fock space describing the classical field using a displacement transformation $D(\beta(t))=\exp (\beta(t) \hat a^\dagger_Y - \beta^* (t) \hat a_Y)$. To determine $\beta(t)$ we collect driven terms linear in $\hat a^\dagger$ from the transformed Hamiltonian, which we set to zero:
\begin{equation}
    (\omega_Y \beta+i\alpha(t)-i\dot\beta)\hat a^\dagger_Y=0.\label{eq:displacement_drop_terms}
\end{equation}
Re-writing to 
\begin{equation}
    \dot \beta = -i\omega_Y\beta + \alpha(t),
\end{equation}
we recognize the equation of motion for a cavity without dissipation. For $\beta(t)$ to fully account for the classical cavity response, including the loss rate $\kappa_Y$, we add the cavity decay term
\begin{equation}
    \dot \beta = -(\frac{\kappa_Y}{2}+i\omega_Y)\beta + \alpha(t).
\end{equation}
We set $\beta(t)$ to the solution of this equation, allowing to drop the terms on the left hand side of Eq. \ref{eq:displacement_drop_terms} from the Hamiltonian:
\begin{multline}
    H = \sum_{i=X,Y}\omega_i \hat a^\dagger_i \hat a_i
    + \sum_{s=\uparrow, \downarrow} \omega_{qd} \ket{T_s}\bra{T_s}
    \\
    + \sum_{i=X,Y}\sum_{s=\uparrow, \downarrow} g \, \hat E_i \cdot \hat d_s \left( \ket{T_s}\bra{s}a_i + \ket{s}\bra{T_s}a^\dagger_i \right)
    \\
    + \tilde \alpha(t) \sum_{s=\uparrow, \downarrow} g_{Y, s} \ket{T_s}\bra{s} 
    + \tilde \alpha^* (t) \sum_{s=\uparrow, \downarrow} g_{Y, s} \ket{s}\bra{T_s}.\label{eq:driven_hamiltonian}   
\end{multline}
The effect of the displacement transformation is moving the drive from the optical mode to the two level system, while correcting for the enhancement of the coupling from the high photon numbers and cavity enhancement of the driving field.
The displacement transformation must be inverted by the transformation $\langle\hat a_Y\rangle \rightarrow \langle\hat a_Y(t)\rangle  + \beta(t)$ when calculating expectation values. During finalization of this work we noticed the recently published Ref. \cite{gonzalez-ruizTwophotonCorrelationsHOM2025} uses a similar displacement transformation.

\subsection{Role of cavity modes}\label{app:swapped_roles}
\begin{figure}
    \centering
    \includegraphics{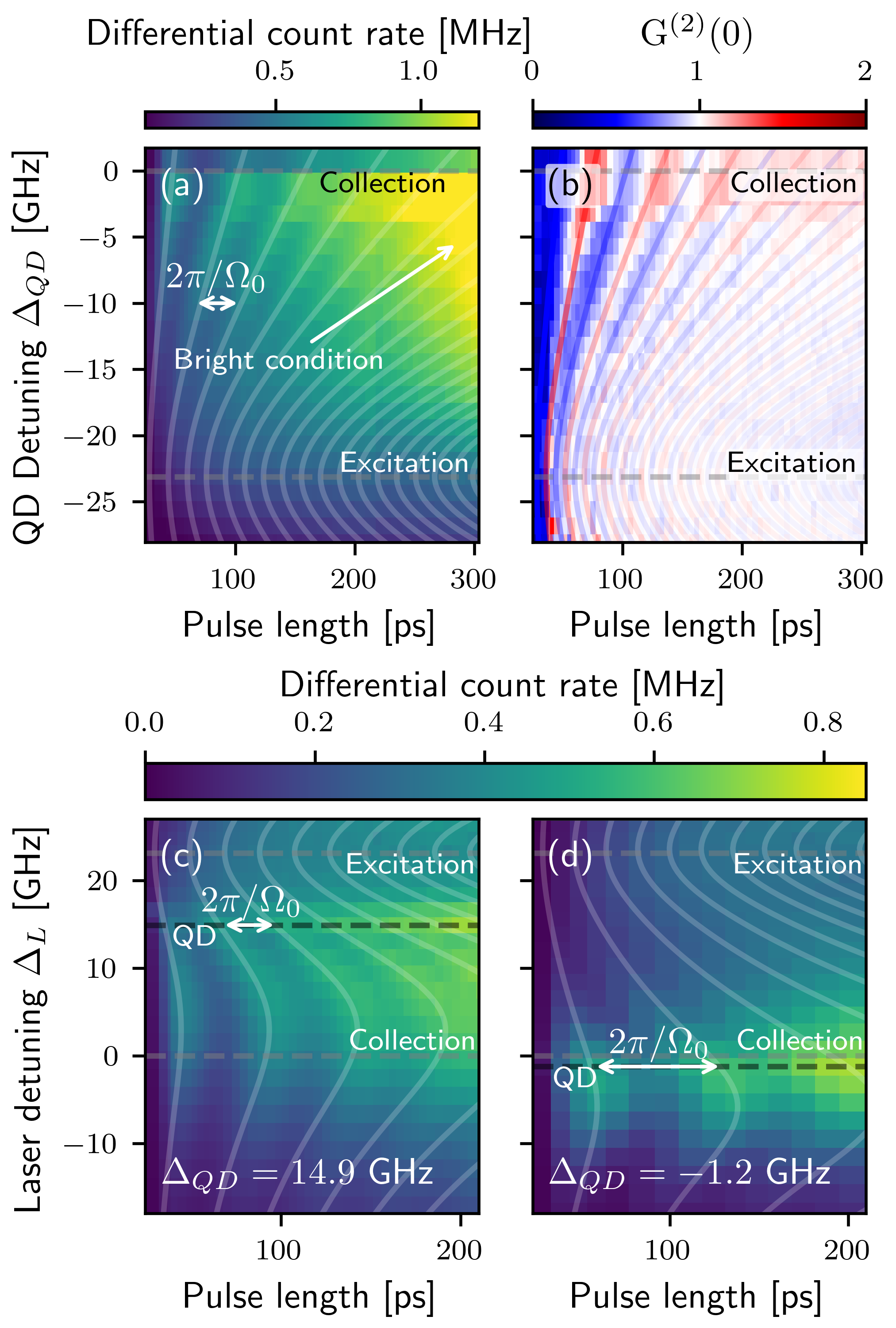}
    \caption{Complementary cavity mode role configuration compared to the main text. (a) Differential count rate where we vary $\Delta_{QD}$ and the pulse length. Here the red (blue) mode is used for excitation (collection), opposite of Fig. \ref{fig:cavity-QD-detuning-red}. (b) Pulsed mode second order correlation function at zero showing oscillating single photons and photon bundles, corresponding to panel (a). (c-d) Sweep of the laser detuning and pulse length with (c) the QD between the collection and excitation modes and (d) the QD in the collection mode. The blue (red) mode is used for excitation (detection), opposite of Fig. \ref{fig:cavity-laser-detuning}.}
    \label{fig:complementary_roles}
\end{figure}
Recent literature investigates the role of the blue and red cavity mode (the respectively higher and lower frequency polarization split modes of the fundamental cavity mode) in resonant excitation and takes into account the effect of the phonon bath \cite{javadiCavityenhancedExcitationQuantum2023, liuUltrafastDepopulationQuantum2016, nazirModellingExcitonPhonon2016}. Figure \ref{fig:complementary_roles} shows the complementary cavity mode roles to the figures in the main text. Here we do not see the associated asymmetries in our data because the peak intensities for the driving pulses used in this work are sufficiently low, avoiding phonon coupling, as explained in the main text.

\bibliography{20260521}

\onecolumngrid

\renewcommand{\thefigure}{S\arabic{figure}}\setcounter{figure}{0}\renewcommand{\theequation}{S\arabic{equation}}\setcounter{equation}{0}\renewcommand{\thetable}{S\arabic{table}}\setcounter{table}{0}

\setcounter{secnumdepth}{2}

\section*{Supplemental Material}
\subsection{Semi-classical model fits}\label{app:semi_classical_model}
\begin{figure}
    \centering
    \includegraphics{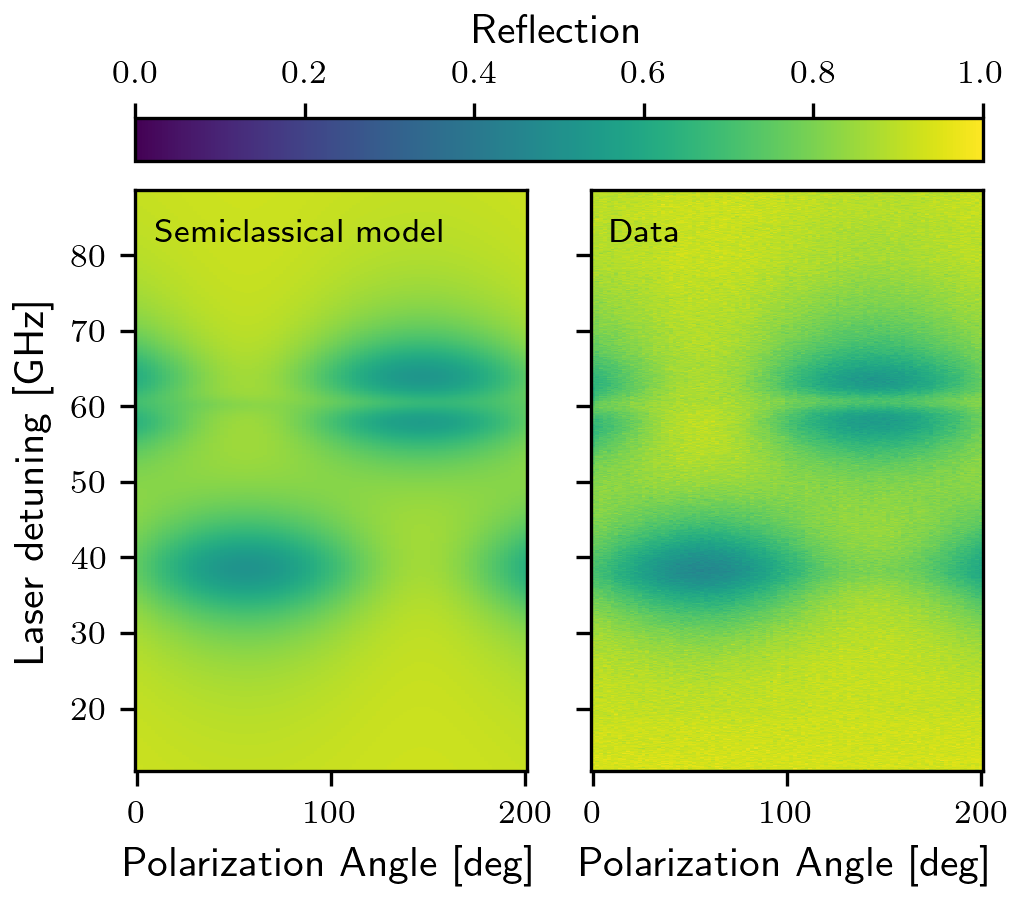}
    \caption{Semiclassical model fit with the QD tuned to the blue mode. The red cavity mode is at a polarization angle of 55 degrees, and the blue cavity mode containing the QD is at 145 degrees. The laser detuning is relative to $320.6$ THz.}
    \label{fig:semiclassical_blue_mode}
\end{figure}

We fit the semi-classical model \cite{snijdersExtendedPolarizedSemiclassical2020} to the reflected counts without polarization filtering as function of the excitation laser frequency and linear polarization angle. An example is shown in Fig. \ref{fig:semiclassical_blue_mode} where the QD is in the blue mode. The fit parameters are the frequencies $f_{red},f_{blue}, f_{QD}$ and the decay rates $\kappa_{red}, \kappa_{blue}, \gamma_0$. Time-resolved measurements show that the pure dephasing rate is small compared to the lifetime, therefore we ignore pure dephasing.

The fit was performed by identifying the system polarization angles, cleaning up global variations in the detected intensity, and fitting the QD and cavity frequencies at these fixed polarization angles by line cuts of the detuning data. Once these rough estimates were found, the full data was used for a fine-tuned fit. 

\subsection{Calibration of detunings}\label{app:calibration_of_detunings}
The cavity splitting and relative frequency of the QD can be determined using the excitation polarization scans used for fitting the semiclassical model, where the excitation laser detuning was calibrated using a wave meter. To find the QD detuning, the QD frequency relative to the cavity modes is assumed to be a linear function of only the QD bias voltage. We obtain a detuning by interpolation of the QD bias voltages that place the transition in each cavity mode. In the experiments where the laser follows the QD resonance, the laser detuning is assumed equal to the QD detuning. We check this assumption by verifying the laser control voltage is linear with the QD bias voltage. However, by comparing with simulations we observe a small systematic offset, caused by a combination of the optimization condition and hysteresis of the laser tuning piezo. In the case that the QD detuning remains fixed, we find the absolute detuning of the laser using the continuous-wave-like feature at the known QD detuning.

\subsection{Pulsed $g^{(2)}(0)$}\label{app:geetwo}
The plots of the second order correlation function at zero time delay are calculated by considering a $400$ ps window around each peak over which coincidence counts are summed. These summed coincidence counts are now normalized such that the average height of the peaks far away from zero have unity magnitude, which we call the pulsed-mode second order autocorrelation function $\mathrm{G}^{(2)}(\tau_k)$ following Ref. \cite{dadaIndistinguishableSinglePhotons2016}, or $\mathrm{G}^{(2)}_k$. We use all peaks except the one at zero delay for normalization because the device does not show significant blinking. 

\subsection{Origin of the measured bunching}\label{app:bunching}
Figure \ref{fig:cavity-laser-detuning}(b) of the main text shows $\mathrm{G}^{(2)}(0) > 1$ below $\Delta_{QD}<3\text{ GHz}$. This is because there, Purcell enhancement is weaker and we collect less fluorescence photons, as shown by Fig. \ref{fig:cavity-QD-detuning-red}(a), and the signal reaching the HBT analyser is dominated by residual excitation light. We have investigated this case, and found that this bunching is an artifact of our pulse generation method which has a small timing jitter ($\sim10\text{ ps}$). This induces a variation of the pulse energy from pulse to pulse, which leads to a measured $\mathrm{G}^{(2)}(0) > 1$. Here, we explain how $\mathrm{G}^{(2)}(0)$ is sensitive to jitter. Then we explain why the observable jitter on the excitation pulse energy does not measurably affect the QD emission.

\begin{figure}
    \centering
    \includegraphics{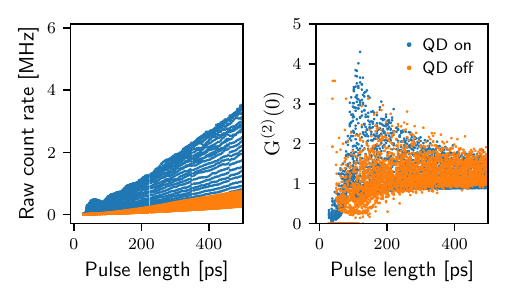}
    \caption{Raw counts and $\mathrm{G}^{(2)}(0)$ with the QD tuned to resonance (QD on) and far out of resonance (QD off). In both cases significant bunching can be observed around 100 ps pulse lengths, showing the bunching originates from the excitation pulses.}
    \label{fig:qd_on_qd_off}
\end{figure}
Figure \ref{fig:qd_on_qd_off} shows that bunching is observed both when the QD is on resonance with the laser and when it is off resonance. Thus, the origin of the bunching is from the excitation laser. By cascading two EOMs the temporal jitter on each pulse changes the pulse energy from pulse to pulse: consider the energy in a `lucky' zero-jitter pulse where the second EOM gate is perfectly synchronous with the pulse from the first, and the energy in an `unlucky' pulse where in the extreme case the second EOM gate opens and closes before (or after) the pulse from the first.

Now we analyze the impact of the pulse jitter on our experiment. First we show the effect of temporal jitter on the pulse area. Each opening of the EOM gate is triggered by a clock. Suppose trigger $i$ has random jitter $\delta t_i$ with $\langle \delta t_i\rangle=0$. The second EOM opens with some delay matched to the optical path length at trigger $n$ to cascade the pulse from trigger $n-m$ generated by the first EOM. The intensity of the resulting pulse is
\begin{equation}
    I_0\Pi_w(t+\delta t_{n-m})\Pi_w(t+\delta t_{n})
\end{equation}
where $\Pi_w(t)$ is a square pulse function of width $w$ and unit height.
The pulse area is (from Eq. \ref{eq:pulse_area}, we require $w>|{\delta t_{n-m} - \delta t_n}|$)
\begin{equation}
\Theta = \frac{\mu \cdot E_0}{\hbar} \left( w - |{\delta t_{n-m} - \delta t_n}| \right).
\end{equation}
The variance of the pulse area jitter scales with the temporal jitter as
\begin{equation}
    \langle{\Theta^2}\rangle-\langle{\Theta}\rangle^2=\left(\frac{\mu \cdot E_0}{\hbar}\right)^2\left\langle{(\delta t_{n-m} - \delta t_n)^2}\right\rangle_n.
\end{equation}
Note that the population inversion at integer multiples of $\pi$ pulse areas is in first order insensitive to pulse area. 

Now consider the effect of jitter on the $k$-th bin of the pulsed-mode second order autocorrelation function of the laser light. Following a similar analysis we calculate
\begin{equation}
    \mathrm{G}^{(2)}_k \approx \frac{w^2 
    - \left\langle\Delta_{n,k,m}^2\right\rangle_{n}}{w^2 
    - \lim_{k\rightarrow \infty}\left\langle\Delta_{n,k,m}^2\right\rangle_{n}},
\end{equation}
under the condition that $\left\langle\Delta_{n,k,m}^2\right\rangle_{n} \ll w$,  where
\begin{equation}
    \Delta_{n,k,m}\equiv\delta t_{n+k} + \delta t_{n+k-m}-  \delta t_n - \delta t_{n-m}.
\end{equation}
This calculation shows that the correlation function is quadratically sensitive to twice as much jitter terms $\delta t_i$, making the second-order correlation function of the driving pulses much more sensitive to pulse jitter than the pulse area. 

Finally we note that our quantum master model shows that using perfect pulses, and shows that the interaction of the QD with the incident laser light also leads to bunching in the reflected state. This can be understood as the one-photon component of the state being used to excite the QD. However this predicted effect is much weaker and at a slightly different detuning than observed .

\subsection{Experimental Protocol}\label{app:experimental_protocol}
The experiment requires various optimizations to either correct slow drifts (EOM bias) or to move to a different experimental condition (placing the laser on the QD). Our experimental protocol for the on-resonance experiments is structured as follows: (1) Set the QD bias voltage. (2) Disable the pulse generation. (3) Optimize the laser detuning to the most detected counts. (4) Sequentially optimize the EOM biases to the least detected counts. (5) Enable the pulse generation and set the pulse length. (6) Measure the coincidence counts, single counts and incident laser power. (7) Detune the QD far away from the collection mode using the QD bias voltage. (8) Measure the detuned coincidence counts, single counts and incident laser power. (9) Set the QD bias voltage back to the value of interest. (10) If the EOM biases were optimized more than five minutes ago, disable pulse generation and optimize the EOM biases. (11) Go to step 5 for the next pulse length value, once all values are used continue. (12) Go to step 1 for the next QD bias value until all values are measured.

Our experimental protocol for the fixed QD detuning is similar as above, but the laser detuning is set in the outer loop instead of using the laser optimization routine.

\subsection{Slow component}\label{app:slow_component}
In Figs. \ref{fig:cavity-laser-detuning} and \ref{fig:complementary_roles}(c) a narrow peak is visible at the QD resonance. To find the origin of this peak, we consider that the counts in this peak go linear with pulse width, which indicates that they are proportional to the number of photons in the pulse thus only weakly driving the QD. We estimate that this driving field is a factor $\sim100$ weaker by comparing this peak height to the linear slope caused by re-excitation in pulsed-mode. The narrowness of the peak further indicates that the driving field pulse length must be much longer than the QD lifetime. 

We rule out multiple optical reflections, because these would be too short and the coincidence counts show no sign of delayed ghost pulses of the appropriate intensity. Genuine continuous-wave leakage through our EOMs is ruled out because the narrow scattering grows linearly with pulse length. 

We suspect that the EOM bias optimization in the experimental protocol (see Experimental Protocol) causes a slow component in the pulse shape. The EOM bias was optimized on the QD fluorescence instead of the driving field power. Later measurements have shown that this can lead to an improper value of the EOM bias causing a slow component in the pulse shape. Similar measurements where the bias was optimized on the laser power have since been performed, these do not show the narrow peak. In the simulations (Fig. \ref{fig:cavity-laser-detuning}(e-f)) we have modeled the slow component by an exponential decay with a characteristic time of 1 ns and an amplitude unique to each panel: the detuning of the QD affects the amplitude of the driving field slow component. This feedback is strong evidence for the proposed origin of the narrow response.

\end{document}